\documentclass[a4paper,twocolumn,11pt]{quantumarticle}
\pdfoutput=1
\usepackage[utf8]{inputenc}
\usepackage[english]{babel}
\usepackage[T1]{fontenc}
\usepackage{float}
\usepackage{amsmath}
\usepackage{amssymb}
\DeclareMathOperator{\ord}{\textit{ord}}
\usepackage{hyperref}
\usepackage{algorithm2e}
\SetKw{KwAnd}{and}
\SetKw{KwOr}{or}
\SetKw{KwTrue}{True}
\SetKw{KwFalse}{False}
\usepackage{tcolorbox}
\usepackage{xcolor}
\definecolor{bggray}{gray}{0.95}
\usepackage[frozencache=true,cachedir=minted-cache]{minted} 
\setminted[python]{
fontsize = \footnotesize,
bgcolor=black!5!white,
}
\usepackage{booktabs}
\usepackage{subcaption}
\usepackage{graphicx}
\usepackage{multirow}
\usepackage{lipsum}

\usepackage{tikz}
\usetikzlibrary{quantikz}
\makeatletter
\DeclareExpandableDocumentCommand{\cwap}{O{}O{1.5pt}O{1.5pt}m}{%optional parameter contains styling info. compulsory is gate text.
	|[inner sep=4pt,minimum width=#2,minimum height=#3]|%
	\edef\n{\the\pgfmatrixcurrentrow} %the row
	\edef\m{\the\pgfmatrixcurrentcolumn} %the column
	\edef\options{row=\n,col=\m,#1}
	\def\DisableMinSize{0}%
	\pgfkeys{/quantikz,wires=2,style=,label style=,braces=,cwires={1,2}}%
	\pgfkeys{/quantikz,#1}%
	\pgfkeysgetvalue{/quantikz/wires}{\quantwires}
	\pgfkeysgetvalue{/quantikz/style}{\a}
	\pgfkeysgetvalue{/quantikz/label style}{\b}
	\pgfkeysgetvalue{/quantikz/cwires}{\mylist}
	\pgfkeysgetvalue{/quantikz/nwires}{\nowires}
	\pgfkeysgetvalue{/quantikz/bundle}{\bundle}
	\def\quantwires{2}
	\phantom{wide}
	\settowidth{\myl}{$wide$}
	\settoheight{\myh}{$wide$}
	\settodepth{\myd}{$wide$}
	\IfInList{1}{\mylist}{\cw}{\IfInList{1}{\nowires}{}{\IfInList{1}{\bundle}{\qwbundle[alternate]{}}{\qw}}}%do we need classical, no wire, or quantum wire?
	\edef\k{\the\numexpr\n+\quantwires-1\relax}
	\edef\mn{\the\numexpr\m-1\relax}
	\ifthenelse{\quantwires=1}{}{%more than 1 wire on gate. iterate through each wire
		\foreach \i in {\the\numexpr\n+1\relax,...,\k} {
			\edef\newcom{\noexpand\vcwhexplicit{\i-\m}{\i-\mn}}
				\edef\newcomb{\noexpand\vqwexplicit{\i-\m}{\i-\mn}}
					\edef\newcomc{\noexpand\vqbundleexplicit{\i-\m}{\i-\mn}}
			\edef\val{\the\numexpr\i+1-\n\relax}
			\IfInList{\val}{\mylist}{\newcom}{\IfInList{\val}{\nowires}{}{\IfInList{\val}{\bundle}{\newcomc}{\newcomb}}}%do we need classical, no wire, or quantum wire?
			\globaldefs=1
			\edef\dotikzset{\noexpand\tikzset{row \i\space column \m/.append style={minimum width={max(\the\myl+8pt,#2)}}}}%
			\dotikzset%
			\edef\undotikzset{\noexpand\tikzset{row \i\space column \m/.style={}}}%
				\expandafter\pgfutil@g@addto@macro\expandafter\tikzcd@atendglobals\expandafter{\undotikzset}%
		}
		\globaldefs=1%
		\edef\dotikzset{\noexpand\tikzset{row \k\space column \m/.append style={minimum height={max(\the\myh+\the\myd+8pt,#3)}}}}%
		\dotikzset%
		\globaldefs=0%
	}
	\expandafter\expandafter\expandafter\expandafter\expandafter\expandafter\expandafter\pgfutil@g@addto@macro\expandafter\expandafter\expandafter\expandafter\expandafter\expandafter\expandafter\tikzcd@atendsavedpaths\expandafter\expandafter\expandafter\expandafter\expandafter\expandafter\expandafter{%
		\expandafter\expandafter\expandafter\expandafter\expandafter\expandafter\expandafter\cswap@end\expandafter\expandafter\expandafter\expandafter\expandafter\expandafter\expandafter{\expandafter\expandafter\expandafter\a\expandafter\expandafter\expandafter}\expandafter\expandafter\expandafter{\expandafter\b\expandafter}\expandafter{\options}{#4}
	}
}

\newcommand{\cswap@end}[4]{
	\def\DisableMinSize{0}
	\pgfkeys{/quantikz,#3}%import options
	\pgfkeysgetvalue{/quantikz/row}{\row}
	\pgfkeysgetvalue{/quantikz/col}{\col}
	\def\quantwires{2}
	\xdef\LoopGG{}
	\foreach \n in {\row,...,\the\numexpr\row+\quantwires-1\relax} {
	\ifnodedefined{\tikzcdmatrixname-\n-\col}{
		\xdef\LoopGG{\LoopGG(\tikzcdmatrixname-\n-\col)}
		}{}
	}
	\node (group\tikzcdmatrixname-\row-\col) [fit=\LoopGG,operator,inner sep=0pt,#1] {\hphantom{Wide}};
	\draw [thickness,transform canvas={yshift=0.05cm}] (group\tikzcdmatrixname-\row-\col.west|-\tikzcdmatrixname-\row-\col.center) to[out=0,in=180] (group\tikzcdmatrixname-\row-\col.east|-\tikzcdmatrixname-\the\numexpr\row+1\relax-\col.center);
	\draw [thickness,transform canvas={yshift=-0.05cm}] (group\tikzcdmatrixname-\row-\col.west|-\tikzcdmatrixname-\row-\col.center) to[out=0,in=180] (group\tikzcdmatrixname-\row-\col.east|-\tikzcdmatrixname-\the\numexpr\row+1\relax-\col.center);
	\draw [line width=3pt,white,shorten >=0.9pt,shorten <=0.9pt] (group\tikzcdmatrixname-\row-\col.east|-\tikzcdmatrixname-\row-\col.center) to[out=180,in=0] (group\tikzcdmatrixname-\row-\col.west|-\tikzcdmatrixname-\the\numexpr\row+1\relax-\col.center);
	\draw [thickness,transform canvas={yshift=0.05cm}] (group\tikzcdmatrixname-\row-\col.east|-\tikzcdmatrixname-\row-\col.center) to[out=180,in=0] (group\tikzcdmatrixname-\row-\col.west|-\tikzcdmatrixname-\the\numexpr\row+1\relax-\col.center);
	\draw [thickness,transform canvas={yshift=-0.05cm}] (group\tikzcdmatrixname-\row-\col.east|-\tikzcdmatrixname-\row-\col.center) to[out=180,in=0] (group\tikzcdmatrixname-\row-\col.west|-\tikzcdmatrixname-\the\numexpr\row+1\relax-\col.center);
}
\makeatother

\DeclareExpandableDocumentCommand{\ctargX}{O{}m}{|[ccrossx2,#1]| {} \cw}
\tikzset{%
	ccrossx/.style={path picture={%
		\draw[internal,inner sep=0pt] (path picture bounding box.south east) -- (path picture bounding box.north west) (path picture bounding box.south west) -- (path picture bounding box.north east);
 }},
	ccrossx2/.style={circle,ccrossx,minimum size=1em},
}

\title{End-to-end compilable implementation of quantum elliptic curve logarithm in Qrisp}

\author{Diego Polimeni}
\affiliation{Alice \& Bob, 53 Bd du Général Martial Valin, 75015 Paris, France}
\affiliation{Lab for Statistical Mechanics of Inference in Large Systems (SMILS), IC, 
École Polytechnique Fédérale de Lausanne (EPFL), CH-1015 Lausanne, Switzerland}

\author{Raphael Seidel}
\affiliation{Fraunhofer Institute for Open Communication Systems}
\orcid{https://orcid.org/0000-0003-3560-9556}

\begin{document}

\maketitle

\begin{abstract}
Elliptic curve cryptography (ECC) is a widely established cryptographic technique, recognized for its effectiveness and reliability across a broad range of applications such as securing telecommunications or safeguarding cryptocurrency wallets. Although being more robust than RSA, ECC is, nevertheless, also threatened by attacks based on Shor's algorithm, which made it a popular field of study in quantum information science. A variety of techniques have been proposed to perform EC arithmetic in quantum devices; however, software support for compiling these algorithms into executables is extremely limited. Within this work, we leverage the Qrisp programming language to realize one of the first fully compilable implementations of EC arithmetic and verify its correctness using Qrisp's built-in sparse matrix simulator.
\end{abstract}

\section{Overview}
Elliptic curve cryptography (ECC) is a widely established standard for public key cryptography \cite{Ullah_2023}. The underlying one-way function for this type of encryption is arithmetic over finite field elliptic curves, which makes this type of encryption vulnerable to attacks based on Shor's algorithm. While the core idea around such an attack is relatively straightforward and similar to Shor's algorithm for factorization, it is the implementation details of elliptic curve arithmetic that bring a variety of complications into this algorithm. A series of publications gave efficient circuits for several subcomponents of the algorithm in order to conduct resource estimation: 
\begin{figure}[ht]
    \centering
    \includegraphics[width = 0.45\textwidth]{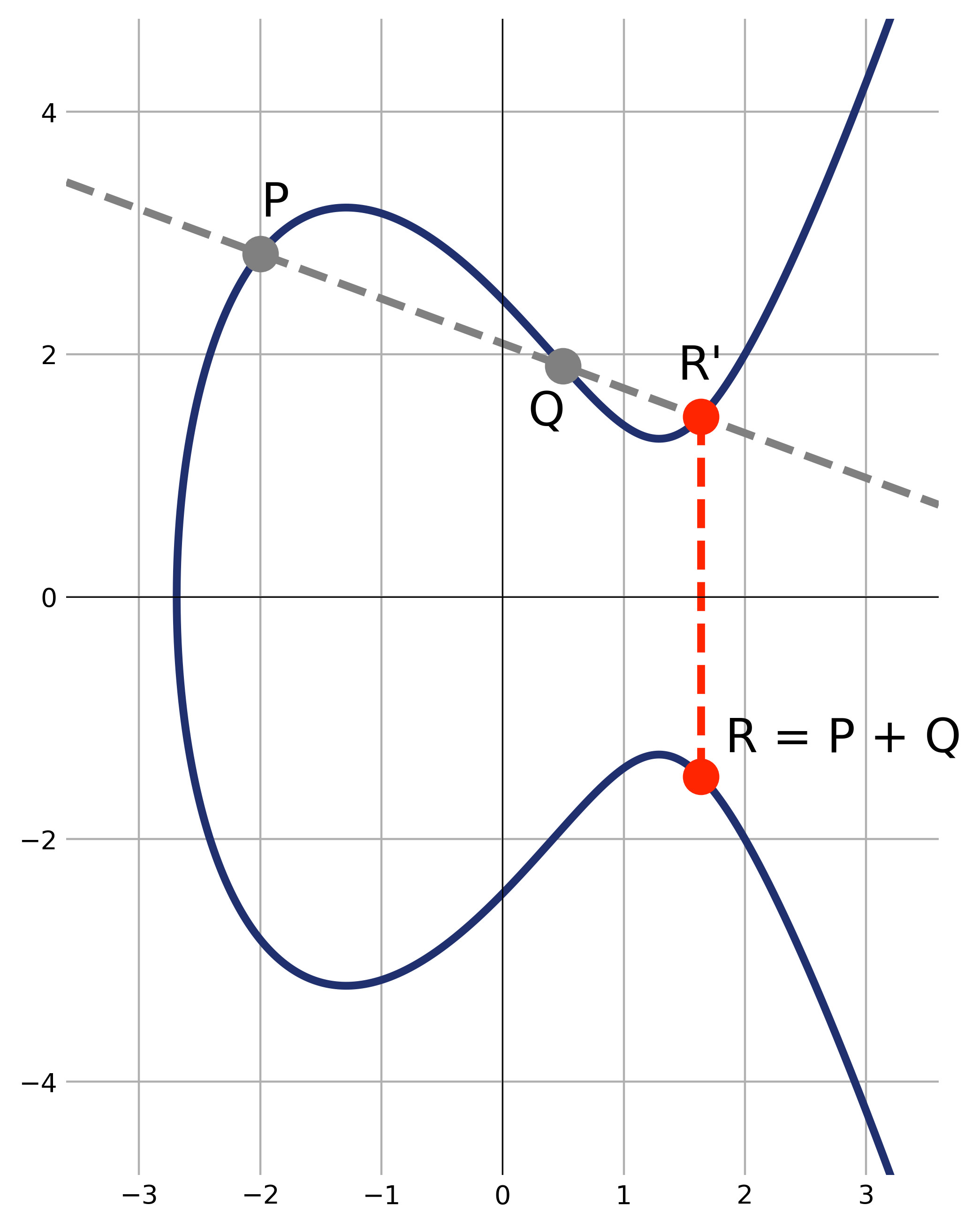}
    \caption{Visualized representation of the addition operation on the elliptic curve $y^2 = x^3 -5x+ 6$ on the real field.}
    \label{fig:Additio_on_cubic}
\end{figure}
Roetteler et al. proposed circuit realizations for controlled elliptic curve addition and estimates are derived from a
simulation in the framework Liqui|$\rangle$ \cite{Roetteler_2017}. Häner et al. improved quantum circuits for elliptic curve scalar multiplication and conducted unit tests and automatic quantum resource estimation through the framework Q\# \cite{Haner_2020} and later Jaques et Häner performed a more detailed study of their implementation by investigating the algorithm through a sparsity-based quantum simulator \cite{jaques_2021}. Gouzien et al. further improved quantum circuits and performed quantum resource estimation on a cat-qubit-based architecture \cite{Gouzien_2023}. Litinski performed an active-volume quantum resource estimation on local and non-local architectures, while introducing three algorithmic modifications that reduce the Toffoli count \cite{litinski_2023}. Until today, however, only one end-to-end implementation, combining the aforementioned techniques into an executable quantum circuit, seems to have emerged. We attribute this to engineering difficulties, specifically because most modern quantum software development still happens by manipulating the assembler-like quantum circuit representation. This problem was recently tackled by a novel quantum programming language called Qrisp \cite{seidel_2024_qrisp}. While Qrisp still supports programming on a low-level, its particular architecture enables seamless usage of low-level program specifications in higher-order routines. The whole implementation of Shor's algorithm for solving the Elliptic Curve Discrete Logarithm Problem (ECDLP) is presented below. It should be noted that the algorithm considered here does not correspond to current state-of-the-art design; improved subroutines can be found in the aforementioned publications. 
\begin{minted}[breaklines]{python}
from qrisp import (QuantumModulus, 
QuantumArray, h, QFT)
import src.classical.ec_arithmetic as clECarithm
import src.quantum.ec_arithmetic as qECarithm
"""
Shor's algorithm to solve 
a small instance of the ECDLP
"""

#Elliptic curve parameters
p=7
a=5
b=4
curve = clECarithm.EllCurve(a, b, p)

# Quantum type for mod p arithmetic
mod_p = QuantumModulus(p)

# Allocate variables holding 
# the elliptic curve point
ecp = QuantumArray(qtype=mod_p, shape=(2,))

#initialize with the point P_0 = [2,6]
ecp[:] = [2,6]

# (Classical) sub-group Generator G = [3,2]
G = [3,2]

# (Classical) Target P = [0,2]
P = [0,2]

# QPE result variables
n = p.bit_length()
x1 = QuantumFloat(n) 
x2 = QuantumFloat(n) 

# Superposition
h(x1)
h(x2)

# In-place EC arithmetic step 1:
# ecp = P0 + x1*G 
qECarithm.ctrl_ell_mult_add(G, ecp, x1, curve)

# In-place EC arithmetic step 2:
# ecp = P0 + x1*G - x2*P
qECarithm.ctrl_ell_mult_add(P, ecp, x2, curve)

# Inverse Quantum Fourier Transform on
# x1 and x2
with inverse():
    QFT(x1)
    QFT(x2)
\end{minted}

\section{Elliptic curve arithmetic}
\paragraph{Elliptic curve definition} 
An elliptic curve over a field $\mathbb{K}$, of characteristic different from 2 and 3, is a projective, non-singular curve of genus 1. It can be defined as the locus of points $(x,y) \in \mathbb{K} \times \mathbb{K}$ satisfying the Weierstrass equation:

\begin{equation}
    E: y^2 = x^3 + ax + b
    \label{weierstrass}
\end{equation}
where $a, b \in \mathbb{K}$ are constants. The curve is projective since it contains the point at infinity, $O$. The curve is non-singular if $4a^3 +27b^2 \neq 0$ which ensures that it doesn't have cusps or nodes. It is relevant to notice that the curve is symmetric with respect to the $x$-axis, property easily verified by applying the transformation $y \mapsto -y$ to Eq. \eqref{weierstrass}. An elliptic curve is best visualized over the real field $\mathbb{R}$, as shown of Fig. \ref{fig:Additio_on_cubic}. Nonetheless, for most cryptographic applications, including the bitcoin digital signature, the field of interest is the finite field $\mathbb{F}_p$ for $p > 3$ a prime number.
The set of points consisting of $O$ and all solutions $(x,y) \in \mathbb{F}_p \times \mathbb{F}_p$ to Eq. \eqref{weierstrass} is denoted by:
\begin{equation}
\begin{split}
    E(\mathbb{F}_p) =
    &\{(x,y) \in \mathbb{F}_p \times \mathbb{F}_p \hspace{0.1cm}|\hspace{0.1cm} y^2 = x^3 + ax + b\}\\ &\cup \{O\}
\end{split}
\end{equation}

The set $E(\mathbb{F}_p)$ is an abelian group with respect to a group operation "$+$", called addition, that is defined via rational functions in the point coordinates with $O$ as the neutral element. An extensive treatment of elliptic curves can be found in \cite{silverman_2009}.

\paragraph{Addition} The geometric definition of the addition over the elliptic curve is the following: let $P$ and $Q$ be two points on the elliptic curve, the line passing through them intersects the curve at most at another point $R'$; the result of the addition $R = P + Q$ is the symmetric of $R'$ with respect to the $x$-axis.
Some particular cases arise and need to be defined separately:
\begin{itemize}
    \item If $P$ and $Q$ are joined by a vertical line, then the result of the addition is the point at infinity: $P + Q= O$.
    \item If $P$ and $Q$ are joined by a line tangent to the curve in $P$ (respectively, $Q$), the intermediate point $R'$ is equal to $P$ (respectively, $Q$), as the result of a double contact between the line and the curve. The result of the addition is then $P + Q = -P$, (respectively, $P + Q = -Q$).
    \item If $P=Q$, then the line that joins them is tangent to the curve in $P$. The intermediate point $R'$ will be the intersection of the tangent with the curve, and the result $R$ will be the symmetric point with respect to the $x$-axis.
\end{itemize}
The geometric definition is consistent with the group structure, since when $Q = O$ the intermediate point will be $-P$ and the result of the addition will be $P$, highlighting the fact that $O$ is the neutral element for the addition over the elliptic curve.

An equivalent algebraic definition of the addition on the elliptic curve can be given by representing the points on the curve by their coordinates: $P = (x_P , y_P )$ and $Q = (x_Q, y_Q)$. When $x_Q \neq x_P$
the slope of the line is given by $\lambda = \frac{y_Q - y_P}{x_Q - x_P} $ division is well-defined as the elliptic curve is defined over a field. The coordinates of the intermediate point $R'$ are obtained by solving the system describing the intersection between the curve and the line, and discarding the solutions corresponding to the points $P$ and $Q$. These coordinates are:
\begin{subequations}
\begin{align}
x_{R'} &= \lambda^2 - x_P - x_Q\\
y_{R'} &= y_P + \lambda(x_{R'} - x_P )
\end{align}
\label{coord}
\end{subequations}
As discussed, the resulting point $R$ will have coordinates:
\begin{subequations}
\begin{align}
x_{R} &= x_{R'} \\
y_{R} &= -y_{R'}
\end{align}
\end{subequations}

When $x_Q = x_P$ and $y_Q \neq y_P$, for the horizontal symmetry of the curve, $y_Q = -y_P$, which means that $Q = -P$. The result of the addition $P + (-P) = O$ is the neutral element. 
When $x_Q = x_P$ and $y_Q = y_P$, which translates to $P = Q$, the slope of the tangent is given by $\lambda = \frac{3x^2_P + a}{2y_P} $ and the coordinates of the result are given by Eq. \eqref{coord}, except in the case where $y_Q = y_P = 0$, for which the result of the addition is the neutral point.

\paragraph{Scalar multiplication}
By exploiting the definition of the addition over the elliptic curve, it is possible to define the multiplication of a point by a scalar integer in $\mathbb{Z}$ as such:
\begin{subequations}
\begin{align}
    kP &:= \underbrace{ P + P + ...  + P}_{k \text{ times}}\\
    0 P &:= O\\
    -k P &:= k(-P)
\end{align}
\end{subequations}
It is easily verified that the multiplication by a scalar integer is:
\begin{itemize}
    \item Associative
    \item Distributive
    \item Compatible with the identity
\end{itemize}
Scalar multiplication (or group exponentiation in the multiplicative setting) is one of the main ingredients for discrete-logarithm-based cryptographic protocols. As such it plays a fundamental role in Shor’s algorithm for solving the Elliptic Curve Discrete Logarithm Problem (ECDLP). An essential concept is the order of a point $P$, noted $\ord(P)$, which is the order of the cyclic subgroup generated by $P$, i.e. the smallest integer $r$ such that $rP = O$.

\section{Description of the algorithm}
Provided the context described in the previous sections, we define an instance of the ECDLP as such: let $G \in E(\mathbb{F}_p)$ be a fixed and classically known generator
of a cyclic subgroup of $ E(\mathbb{F}_p)$ of known order $ord(G) = r$, let $P \in \langle G \rangle$ be a fixed and classically known element in the subgroup generated by $G$; the problem is to retrieve the unique integer $l \in \{1, . . . , r\}$, called the discrete logarithm, such that $P = lG$.
Shor’s algorithm \cite{Shor_1997}, which provides a way to efficiently find the discrete logarithm, proceeds as follows:
\begin{itemize}
    \item Prepare two registers in a superposition of all possible integers between 0 and a large number.
    \item Apply a periodic function which takes as input the first two registers and compute the output in an auxiliary register.
    \item Apply an inverse Quantum Fourier Transform to reveal the period of the function.
\end{itemize}

Shor's algorithm for solving the ECDLP is analogous to Shor's algorithm to solve the factoring problem, as it prepares a large superpositions of inputs, feeds them to a periodic function and exploits the QFT routine to reveal the hidden period, which, in turn, allows to efficiently compute the logarithm. The function mentioned in point two is:

\begin{equation}
f\left(x_{1}, x_{2}\right)=x_{1} G-x_{2} P =\left(x_{1}-x_{2} l\right) G
\end{equation}
where the last equality holds for the definition of the discrete logarithm $l$.
The function $f$ is periodic in both variables, in the following fashion:

\begin{equation}
\forall k, f\left(x_{1}+k l, x_{2}+k\right)=f\left(x_{1}, x_{2}\right)
\end{equation}

Following the outline proposed above, two registers, $\left|x_{1}\right\rangle$ and $\left|x_{2}\right\rangle$ are initialized in a superposition of all possible numbers between 0 and $r-1$:

\begin{equation}
\frac{1}{r} \sum_{x_{1}=0}^{r-1} \sum_{x_{2}=0}^{r-1}\left|x_{1}\right\rangle\left|x_{2}\right\rangle
\end{equation}

The function $f$ is evaluated on those registers as input, and the output is stored in a new register:

\begin{equation}
\frac{1}{r} \sum_{x_{1}=0}^{r-1} \sum_{x_{2}=0}^{r-1}\left|x_{1}\right\rangle\left|x_{2}\right\rangle\left|f\left(x_{1}, x_{2}\right)\right\rangle
\end{equation}

Then, the inverse quantum Fourier transform routine is applied to the registers $\left|x_{1}\right\rangle$ and $\left|x_{2}\right\rangle$. The operation yields the following result:

\begin{equation}
\frac{1}{r^{2}} \sum_{x_{1}, x_{2}, y_{1}, y_{2}=0}^{r-1} e^{2 \pi i\left(x_{1} y_{1}+x_{2} y_{2}\right) / r}\left|y_{1}\right\rangle\left|y_{2}\right\rangle\left|f\left(x_{1}, x_{2}\right)\right\rangle
\end{equation}

We rewrite the sum for clearer readability as such:
\begin{widetext}
\begin{equation}
\sum_{y_{1}, y_{2}=0}^{r-1} \sum_{k=0}^{r-1}\left[\frac{1}{r^{2}} \sum_{\substack{x_{1}, x_{2}=0 \\ f\left(x_{1}, x_{2}\right)=k G}}^{r-1} e^{2 \pi i\left(x_{1} y_{1}+x_{2} y_{2}\right) / r}\right]\left|y_{1}\right\rangle\left|y_{2}\right\rangle|k G\rangle
\label{probamp}
\end{equation}
\end{widetext}

In this form, it is apparent that the state $\left|y_{1}\right\rangle\left|y_{2}\right\rangle|k G\rangle$ has a probability amplitude given by the inner expression of the sum. The sum can be further rewritten noticing that the condition $f\left(x_{1}, x_{2}\right) = \left(x_{1}-x_{2} l\right) G=k G$ leads to $x_{1}-x_{2} l=k \bmod r$, that is, $x_{1}=k+x_{2} l \bmod r$. This last equality, plugged in Eq. \eqref{probamp}, yields the following probability amplitudes associated with the states $\left|y_{1}\right\rangle\left|y_{2}\right\rangle|k G\rangle$:

\begin{equation}
\frac{1}{r^{2}} \sum_{x_{2}=0}^{r-1} e^{2 \pi i\left(k y_{1}+\left(y_{2}+l y_{1}\right) x_{2}\right) / r}
\end{equation}

If $y_{2} + l y_{1}=0 \bmod r$, the exponential has zero argument, hence it will contribute to the sum.
In any other case, the sum vanishes, as effect of destructive interference; this can be seen by developing the sum as a geometric sum and noticing that the numerator is always zero, since the argument of the exponential is always a integer multiple of $2\pi i$.
Finally, from the measure of the registers encoding $y_{1}$ and $y_{2}$, the discrete logarithm is retrieved using $l=-y_{2} y_{1}^{-1} \bmod r$, as long as $y_{1} \neq 0$. It should be noted that both $y_{2}$ and $k$ take $r$ different values, which means that the measurement procedure will yield $r^2$ different results, with uniform probability $\frac{1}{r^2}$. However, $r$ of these will lead to $y_{2}=0$ and $y_{1}=0$, thus they are not useful to find the logarithm. Hence, the probability of obtaining the value of the discrete logarithm is $1-\frac{1}{r}$.
\subsection{Montgomery form}
The Montgomery Form technique \cite{Montgomery_85} is a alternative representation for numbers that operate under modular arithmetic. The motivation to use such a technique is the fact that the divisions required for modular reduction are expensive on most computing hardware and should therefore be avoided.\\
For an arbitrary modular number $a \in \mathbb{F}_p$, the Montgomery form is defined as 
\begin{align}
    \tilde{a} = (R a) \text{ mod } p
\end{align}
where $R$ is an arbitrary number with $\gcd(R, p) = 1$. Modular reduction on numbers in Montgomery form can be achieved through an algorithm called Montgomery reduction.
Although Montgomery reduction still requires divisions, they can be performed with respect to the divisor $R$. This can be chosen to be a power of two, thus replacing expensive division with cheap bit shifts. 
It should be noticed that addition and subtraction in Montgomery form is analogous to standard modular addition and subtraction: 
\begin{subequations}
\begin{align}
    a2^n + b2^n &= (a+b)2^n\\
    a2^n - b2^n &= (a-b)2^n
\end{align}
\end{subequations}
Multiplication in Montgomery form is slightly less straight-forward. In fact, multiplying $a2^n$ and $b2^n$ does not produce the product of a and b in Montgomery form because it has an extra factor of $2^n$:
\begin{widetext}
\begin{equation}
        (a2^n\bmod p) ( b 2^n \bmod p) \bmod p = ( a b 2^n ) 2^n \bmod p
\end{equation}
\end{widetext}

To obtain the actual result of the multiplication in Montgomery form, the extra factor of $2^n$ should be removed. This is simply achieved by multiplying by the modular inverse of $2^n$, conveniently noted $2^{-n}$. 
The multiplication of two \mintinline{python}{QuantumModulus} has been implemented using the ideas described in \cite{rines_2018} and yields a result with a non-trivial Montgomery shift. Arithmetic operations between \mintinline{python}{QuantumModulus} with differing Montgomery shifts could be problematic. There exists two ways around that: the first one is to revert the \mintinline{python}{QuantumModulus} back to standard representation, using the following functions. 
\begin{minted}[breaklines]{python}
def to_montgomery_qm(x: QuantumModulus, montgomery_shift: int):
    x *= pow(2, montgomery_shift, x.modulus)
    x.m = montgomery_shift

def to_standard_qm(x: QuantumModulus):
    montgomery_shift = x.m
    x *= pow(2, -montgomery_shift, x.modulus)
    x.m = 0
\end{minted}
The second one is needed to implement the arithmetic between \mintinline{python}{QuantumModulus} with different Montgomery shifts, commented on in section \ref{sec:mont_form_adder}.
\subsection{Kaliski's algorithm}
\begin{figure*}
\resizebox{0.95\linewidth}{!}{
	\begin{quantikz}[column sep=1em,row sep=1em]
		\lstick{$\ket{u}$} &\qwbundle{n}&\qw &\qw\slice{1}&\octrl{5}&\qw &\qw &\qw\slice{2}&\gate[2,style={rounded corners}]{u > v}\vqw{4}\slice{3}&\swap{1} &\qw\slice{4}&\gate[style={rounded corners}]{\text{Input }u}\vqw{1}&\qw\slice{5} &\qw &\qw &\qw &\qw &\qw &\swap{1} &\qw &\rstick{$\ket{u}$} \qw \\
		\lstick{$\ket{v}$} &\qwbundle{n}&\gate[style={rounded corners}]{=0}\vqw{5}&\qw &\qw &\octrl{4}&\qw &\qw & &\targX{} &\qw &\gate{-u} &\qw &\qw &\qw &\gate{/2}&\qw &\qw &\targX{} &\qw &\rstick{$\ket{v}$} \qw \\
		\lstick{$\ket{r}$} &\qwbundle{n}&\qw &\qw &\qw &\qw &\qw &\qw &\qw &\qw &\swap{1} &\qw &\gate[style={rounded corners}]{\text{Input }r}\vqw{1}&\qw &\qw &\qw &\gate{\times 2 \mod p}&\swap{1} &\qw &\qw &\rstick{$\ket{r}$} \qw \\
		\lstick{$\ket{s}$} &\qwbundle{n}&\qw &\qw &\qw &\qw &\qw &\qw &\qw &\qw &\targX{} &\qw &\gate{+r} &\qw &\qw &\qw &\qw &\targX{} &\qw &\octrl{2}&\rstick{$\ket{s}$} \qw \\
		\lstick{$\ket{b=0}$} &\qw &\qw &\qw &\qw &\qw &\targ{} &\targ{} &\octrl{2} &\qw &\qw &\octrl{-3} &\octrl{-1} &\targ{} &\targ{} &\qw &\qw &\qw &\qw &\qw &\rstick{$\ket{0}$} \qw \\
		\lstick{$\ket{a=0}$} &\qw &\qw &\qw &\targ{} &\octrl{1}&\ctrl{-1}&\qw &\targ{} &\ctrl{-5}&\ctrl{-3} &\qw &\qw &\qw &\ctrl{-1}&\qw &\qw &\ctrl{-3}&\ctrl{-5}&\targ{} &\rstick{$\ket{0}$} \qw \\
		\lstick{$\ket{m_i=0}$}&\qw &\targ{} &\ctrl{1} &\qw &\targ{} &\qw &\ctrl{-2} &\targ{} &\qw &\qw &\qw &\qw &\ctrl{-2}&\qw &\qw &\qw &\qw &\qw &\qw &\rstick{$\ket{m_i}$}\qw \\
		\lstick{$\ket{f}$} &\qw &\ctrl{-1} &\targ{} &\ctrl{-2}&\ctrl{-1}&\qw &\qw &\ctrl{-1} &\qw &\qw &\ctrl{-3} &\ctrl{-3} &\qw &\qw &\ctrl{-6}&\qw &\qw &\qw &\qw &\rstick{$\ket{f}$} \qw
\end{quantikz}
}
    \caption{Quantum circuit for the Montgomery-Kaliski round function, which corresponds to one iteration of the loop, taken from \cite{Gouzien_2023}. Gates controlled by a register are to be understood as to be controlled by the parity of the integer encoded in the register, so the state of the least significant bit. Gates controlled by white dots are triggered when the control qubit is in the '0' state.}
    \label{fig:KaliskiIteration}
\end{figure*}
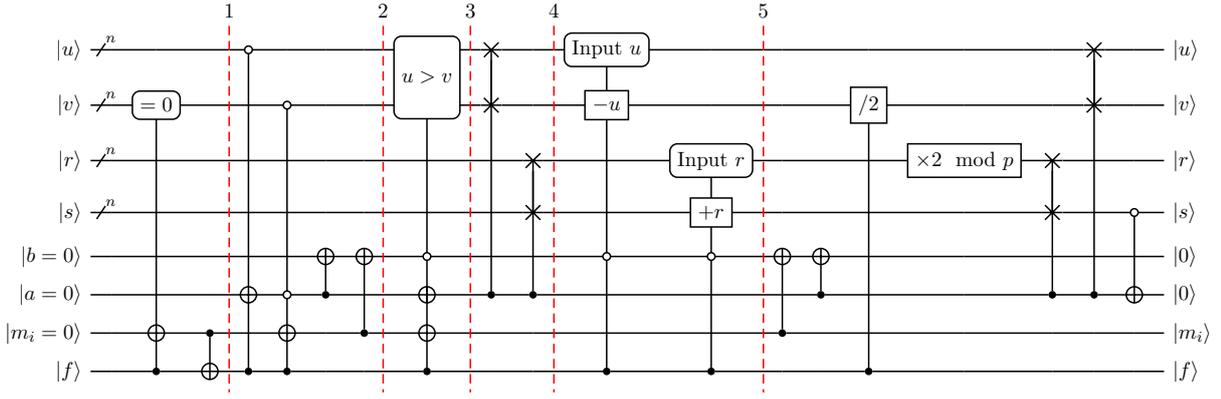
Qrisp offers a set of elementary tools for modular arithmetic, which is, however, not yet complete. A fundamental ingredient currently missing is the modular inversion, used, in conjunction with modular multiplication, to perform modular division and compute slope $\lambda$. The modular multiplicative inverse $x^{-1}$ of a number $x$ is such that $xx^{-1} = x^{-1}x = 1 \pmod{p}$. First, since the Montgomery form is employed, the modular inversion algorithm must be compatible with it. Kaliski's algorithm is a classical algorithm for modular inversion, compatible with Montgomery form, which can be adapted to handle superpositions of inputs. The interesting aspect of the quantum version of the algorithm is that no additional entanglement is generated within it.
\RestyleAlgo{ruled}

\begin{algorithm}[]
\caption{Kaliski's algorithm with swaps \cite{Gouzien_2023, kaliski_95}}\label{alg:kaliski}
$u\gets p$\;
$r \gets 0$\;
$s \gets 1$\;
\For{$i\gets 0$ \KwTo $2*n$}{
    \If{v = 0}{
    $r \gets 2*r$\;
    }
    $swap \gets \KwFalse$\;
    \If{(u is even \KwAnd v is odd) \KwOr \\ \quad(u is odd \KwAnd v is odd \KwAnd u$>$v)}{
        $u \longleftrightarrow v$\;
        $r \longleftrightarrow s$\;
        $swap \gets \KwTrue$\;
    }
    \If{u is odd \KwAnd v is odd}{
        $v \gets v-u$\;
        $s \gets s+r$\;
    }
    $v \gets v//2$\;
    $r \gets 2*r$\;
    \If{r $\geq$ p}{
        $r \gets r-p$\;
    }
    \If{swap is \KwTrue}{
        $u \longleftrightarrow v$\;
        $r \longleftrightarrow s$\;
    }
    $r \gets r-p$\;
    $u \longleftrightarrow r$\;
}
\end{algorithm}
As presented in \cite{kaliski_95}, the algorithm initializes the variables $u = p$, $r = 0$ and $s = 1$. $v < p$ is the number to invert, which is given in Montgomery form. The variables have the final value of $u = 1$, $ s = p$,  and $r$ such that $p-r = v^{-1} 2^{2n} \pmod{p}$ at the end of the iterations. The modular inverse of $v$ is stored in the register containing $v$ itself, making it an in-place operation. A more complete treatment of the state of the variables can be found in \cite{Gouzien_2023}. The pseudocode, which combines two different versions of the Kaliski algorithm, is shown below:

Implementing a high-level version of the algorithm, following the pseudocode, revealed to be feasible, and classical inputs were correctly treated; however, compilation time and simulation time for superposition of inputs were considerably high. To maintain the scalability of our implementation, we disqualified this approach to be used as a subroutine and instead developed a more low-level approach, restricted to more primitive operations such as simple Toffoli or Fredkin gates. For the comparisons and the modular arithmetic, we used the Qrisp defaults.
Fig. \ref{fig:KaliskiIteration}, taken from \cite{Gouzien_2023}, represents the circuit for one iteration of the loop.
Notice that the algorithm would stop when the condition $v = 0$ is reached, which happens between $n$ and $2n$ iterations of the loop; however, to ensure reversibility of the quantum circuit and handle worst cases, the unnecessary operations are deactivated and the remaining iterations of the loop are run anyway.
The last qubit, in state $|f\rangle$, modelled through \mintinline{python}{QuantumBool}, signals when the "stop" condition $v=0$ is met. Initially set to $1$, it changes to $0$ at the start of the first iteration where $v=0$ is detected. After this, only modular doubling of $r$ and uncomputations are performed.
Qubits $|a\rangle$,  $|b\rangle$ and $|m_i\rangle$ are used to decide which branch of the algorithm is run and are modelled through \mintinline{python}{QuantumBool}. After the comparison step, the variable $a$ models the condition for applying the swaps in the pseudocode (second \verb|if|), while the logical complement of $b$ models
the condition for applying the subtraction and addition in the pseudocode (third \verb|if|). After the modular doubling and before the controlled swap operation, a modular reduction of r is performed; this operation is implicitly grouped with the doubling, on the circuit.
It is important to notice that while $a$ and $b$ are allocated within the algorithm and can therefore be deallocated within it, leveraging some conserved quantities, a garbage qubit $|m_i\rangle$ is produced for each iteration of the algorithm and remains polluted after the algorithm. For this reason, a \mintinline{python}{QuantumArray} of preallocated \mintinline{python}{QuantumBool} is passed as an argument to the function. The uncomputation strategy is discussed in the next section.
Comparisons and arithmetic operations are not detailed here since they are built-in functions in Qrisp.

The whole Kaliski's algorithm is implemented by the following function:

\begin{minted}[breaklines]{python}
from qrisp import (QuantumBool, QuantumFloat, 
QuantumModulus, QuantumArray, control, invert,
x, cx, mcx, swap, cyclic_shift)

def kaliski_quantum(v: QuantumModulus, 
                    m: QuantumArray):
    """
    Utilizes the Kaliski-Algorithm to perform 
    an in-place mod inversion on the 
    QuantumModulus v.
    
    m is a QuantumArray cotaining pre-allocated 
    QuantumBools, which are required for the 
    algorithm as temporary values but will be 
    uncomputed using Bennet's trick.
    """

    p = v.modulus
    n = p.bit_length()
    # Convert to Montgomery Form
    to_montgomery(v, p)
    u = QuantumFloat(n)
    u[:] = p
    r = QuantumModulus(2 * p)
    r[:] = 0
    s = QuantumModulus(2 * p)
    s[:] = 1

    a = QuantumBool()
    b = QuantumBool()
    add = QuantumBool()
    f = QuantumBool()
    f[:] = True
    for i in range(2 * n):
        is_zero = v == 0
        mcx([f, is_zero], m[i])
        is_zero.uncompute()
        cx(m[i], f)
        # STEP 1
        mcx([f, u[0]], a, ctrl_state="10")
        mcx([f, a, v[0]], m[i], ctrl_state="100")
        cx(a, b)
        cx(m[i], b)

        # STEP 2
        l = u > v
        mcx([f, l, b], a, ctrl_state="110")
        mcx([f, l, b], m[i], ctrl_state="110")
        l.uncompute()

        # STEP 3
        with control(a):
            swap(u, v)
            swap(r, s)

        # STEP 4
        mcx([f, b], add, ctrl_state="10")
        with control(add):
            v -= u
            s += r
        # STEP 5
        mcx([f, b], add, ctrl_state="10")
        # uncompute b
        cx(m[i], b)
        cx(a, b)

        # Division by 2
        with control(f):
            with invert():
                cyclic_shift(v)

        cyclic_shift(r)
        larger = r > p
        with larger:
            r -= p
        cx(r[0], larger)
        larger.delete()

        with control(a):
            swap(u, v)
            swap(r, s)
        # uncompute a
        mcx([s[0]], a, ctrl_state="0")

    a.delete()
    add.delete()
    b.delete()

    inpl_rsub(r, p)

    for i in range(v.size):
        swap(v[i], r[i])

    # Uncompute u,s,f
    f.delete()
    x(u[0])
    u.delete()
    r.delete()
    s -= p
    s.delete()
    # Convert back to standard representation
    to_standard(v, p)
    return v
\end{minted}

After $2n$ iterations of the Kaliski's loop, the transformation $|r\rangle \rightarrow |p-r\rangle$ is applied using the \texttt{inpl\_rsub} function, allowing an in-place operation that does not pollute additional qubits.
Using the properties of the two-complement representation, a negation of the bits induces the transformation $|r\rangle \rightarrow |- (r+1) \rangle$. Subsequently we add the modulus + 1 to bring the variable into a positive state, which now represents $|(-r) \text{mod} N \rangle$. The procedure is finalized by performing a semi-classical modular in-place addition with $p$.
This is implemented by the following function:
\begin{minted}[breaklines]{python}
def inpl_rsub(r: QuantumModulus,
              p: int):
    """
    Performs a modular in-place
    right-subtraction.
    Implying after executing this function, the 
    new r has the value (p-r)
    """

    # Use the two complement for negation:
    # NOT(r) = -(r + 1)
    x(r)

    # Make positive again
    # The inpl_adder attribute calls a 
    # user-specified, 2^n modular adder
    r.inpl_adder(r.modulus + 1, r)

    # Modular addition of p
    r += p
\end{minted}

\subsection{Elliptic curve addition}
Addition over elliptic curves, as described in Section 2.1, is implemented in the following manner. First, we perform the classical doubling of the generator $P$; then, we perform the quantum addition of a classically known point to a quantum point. Combining the former and the controlled version of the latter, we obtain the desired operation. 
\begin{figure*}
\resizebox{0.95\linewidth}{!}{
	\begin{quantikz}[column sep=1em,row sep=1em]
		\lstick{$\ket{x_1}$}  &\qwbundle{n} &\gate{-x_2} \slice{1} &\gate[style={rounded corners}]{inv}\vqw{3} &\qw &\gate[style={rounded corners}]{mul}\vqw{4} &\gate[style={rounded corners}]{inv}\vqw{3} \slice{2} &\qw &\gate{-} &\gate{+3x_2} \slice{3} &\gate[style={rounded corners}]{mul} \vqw{4} \slice{4} &\gate[style={rounded corners}]{inv}\vqw{3} &\qw &\gate[style={rounded corners}]{inv}\slice{5} \vqw{3} &\gate{neg} \slice{6} &\gate{+x_2}&\rstick{$\ket{x_3}$}  \qw \\
		\lstick{$\ket{ctrl}$} &\qw &\ctrl{1} &\qw &\qw &\qw &\qw &\qw &\ctrl{-1}\vqw{2} &\ctrl{-1} &\qw &\qw &\qw &\qw &\ctrl{-1} &\ctrl{1}&\rstick{$\ket{ctrl}$} \qw \\
		\lstick{$\ket{y_1}$} &\qwbundle{n} &\gate{-y_2} &\qw &\gate[style={rounded corners}]{mul}\vqw{2} &\gate{mul} &\qw &\qw &\qw &\qw &\gate{mul} &\qw &\gate[style={rounded corners}]{mul}\vqw{2} &\qw &\qw &\gate{-y_2}  &\rstick{$\ket{y_3}$} \qw \\
		\lstick{$\ket{t_0 = 0}$} &\qwbundle{n} &\qw  &\gate{inv} &\gate[style={rounded corners}]{mul} &\qw &\gate{inv} &\gate{sq}\vqw{1}  &\gate[style={rounded corners}]{-} &\gate{sq} \vqw{1} &\qw &\gate{inv} &\gate[style={rounded corners}]{mul} &\gate{inv} &\qw &\qw   &\rstick{$\ket{0}$} \qw \\
		\lstick{$\ket{\lambda = 0}$} &\qwbundle{n} &\qw &\qw &\gate{mul} &\gate[style={rounded corners}]{mul} &\qw &\gate[style={rounded corners}]{sq} &\qw &\gate[style={rounded corners}]{sq} &\gate[style={rounded corners}]{mul} &\qw &\gate{mul} &\qw &\qw &\qw  &\rstick{$\ket{0}$} \qw 
\end{quantikz}
}
    \caption{Quantum circuit for controlled elliptic curve point addition, adapted from \cite{Roetteler_2017}. Five register are involved, two of which are uncomputed at the end of the operation: two quantum registers, $|x_1\rangle$ and $|y_1\rangle$, to encode the coordinates of the point, a control qubit $|\text{\textit{ctrl}}\rangle$ and two auxiliary registers $|t_0\rangle$ and $|\lambda\rangle$. The rectangles specify the type of operation performed and the registers involved: rounded-corner rectangles signify the input registers while right-corner rectangles signify the output registers.}
    \label{fig:addition}
\end{figure*}
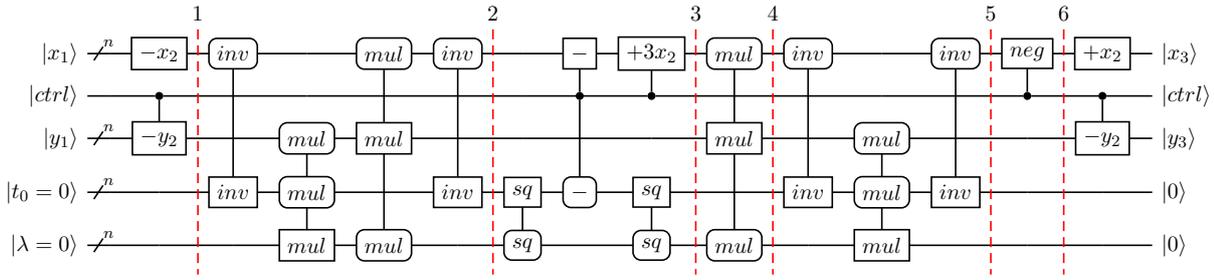
To compute the scalar multiplication $kP$ of a known base point $P$ efficiently, we precompute all the $n$ $2$ -power multiples of $P$ classically. Then we compute the scalar multiple using a sequence of $n$ controlled additions of these precomputed points and store the result in a quantum register, following the binary representation of the scalar. Let $k = \sum_{i=0}^{n-1}k_i2^i$, with $k_i \in \{0,1\}$, be the binary representation of the $n$ bit scalar k. Then:
\begin{equation}
    kP = \sum_{i=0}^{n-1}k_i2^iP = \sum_{i=0}^{n-1}k_i(2^iP)
    \label{binctrlsum}
\end{equation}
This approach offers the benefit that all doubling operations can be performed on a classical computer, leaving the quantum circuit responsible only for the generic point addition. This simplification makes the overall implementation more straightforward and less demanding in terms of resources. The doubling of the classically known generator is implemented as follows:
\begin{minted}[breaklines]{python}
def ell_double(P: list, curve: EllCurve):

    #Return 2P
    p = curve.p
    s = ((3 * (P[0] * P[0] % p) + curve.a) % p) * pow((2 * P[1]) % p, -1, p)
    xr = (s * s - 2 * P[0]) % p
    yr = P[1] - s * ((P[0] - xr) % p) % p

    return [xr, (p - yr) % p]
\end{minted}
Following \cite{Roetteler_2017} and \cite{Gouzien_2023}, the implementation of elliptic curve addition focuses on the generic case where the two points are distinct, not inverses of each other, and neither is the neutral element. These exceptional cases are rare, except when the accumulation register is initialized to the neutral element. To avoid this issue, the accumulation register should be initialized with a point other than the neutral element; this trick does not affect the measurement statistics after the Fourier transform, since it only adds a global phase to the final state. As described above, a classically known point is added to a point whose coordinates are encoded in a quantum register.
The implementation of the addition in Qrisp follows the circuit given on Fig. \ref{fig:addition}, adapted from \cite{Roetteler_2017}, to which we refer the reader for more details.
The addition is implemented as in-place operation, meaning that the coordinates of the result are stored in the registers containing the coordinates of the inputs. 
\begin{minted}[breaklines]{python}
from qrisp import (custom_control, QuantumArray, 
QuantumBool, QuantumModulus, conjugate, 
control, cx, Qubit)

@custom_control
def ell_add_inpl(anc: QuantumArray,
                       G: list, 
                       p: int, 
                       ctrl: Qubit = None):
    """
    Performs in-place elliptic curve point 
    addition of a point whose coordinates are 
    stored in a quantum register and a 
    classically known point; the result of the 
    operation is stored in the registers
    initially containing the coordinates of the
    input.
    """
    # STEP 1
    if ctrl is None:
        anc[1] -= G[1]
    else:
        with control(ctrl):
            anc[1] -= G[1]
    anc[0] -= G[0]

    # STEP 2
    l = QuantumModulus(p)

    # m is an array of temporary QuantumBools, 
    # which will be uncomputed Bennett style
    m = QuantumArray(qtype=QuantumBool(), 
                     shape=2*p.bit_length())
    
    with conjugate(kaliski_quantum)(anc[0], m) as inv:
        temp = anc[1] * inv
        to_standard_qm(temp)
        l[:] = temp
        temp.uncompute()
    for a in m:
        a.delete()

    temp = l * anc[0]
    to_standard_qm(temp)
    anc[1] -= temp
    temp.uncompute()

    # STEP 3
    if ctrl is None:
        anc[0] += 3 * G[0]
    else:
        with control(ctrl):
            anc[0] += 3 * G[0] 

    temp = l * l 
    to_standard_qm(temp)
    if ctrl is None:
        anc[0] -= temp
    else:
        with control(ctrl):
            anc[0] -= temp
            
    temp.uncompute()

    # STEP 4
    temp = l * anc[0]
    to_standard_qm(temp)
    anc[1] += temp
    temp.uncompute()

    # STEP 5
    m = QuantumArray(qtype=QuantumBool(),                          shape=2 * p.bit_length())

    with conjugate(kaliski_quantum)(anc[0], m) as inv:
        temp = anc[1] * inv
        to_standard_qm(temp)
        cx(temp, l)
        temp.uncompute()

    for a in m:
        a.delete()

    # STEP 6
    l.delete()
    if ctrl is None:
        anc[1] -= G[1]
    else:
        with control(ctrl):
            anc[1] -= G[1] 

    # STEP 7
    anc[0] -= G[0]

    if ctrl is None:
        inpl_rsub(anc[0], p)
    else:
        with control(ctrl):
            inpl_rsub(anc[0], p)

    return anc    
\end{minted}
Finally, the classical and quantum routines are combined together, using the decomposition provided by Eq. \ref{binctrlsum} to control the addition. The controlled version of the addition is much more expensive than the plain version, which leads us to implement a series of optimizations as the \mintinline{python}{custom_control} decorator, whose functioning is detailed in \cite{seidel_2024_qrisp}. 
\begin{minted}[breaklines]{python}
from qrisp import control, invert, cyclic_shift

def ctrl_ell_mult_add(power: list, 
                       res: QuantumArray,
                       k: QuantumFloat,
                       curve: EllCurve):
    # Elliptic curve multiplication Q + kP
    n = k.size
    p = curve.p
    for i in range(n):
        with control(k[i]):
            res = ell_add_inpl(res, power, p)
        power = ell_double(power, curve)
    return res
\end{minted}

\subsection{Modular adder from an arbitrary non-modular adder}
As elaborated, the EC addition requires more primitive arithmetic operations such as a modular adder. Such an adder has been described in \cite{beauregard_2003}; however, it uses QFT based addition, which is expected to be particularly costly on fault tolerant devices due to the cost of synthesizing arbitrary angle rotations \cite{ross_2016}. We generalize the overall strategy to enable the use of an arbitrary (non-modular) adder. For instance, it would be possible to leverage Gidney's adder \cite{Gidney_2018} to perform modular addition using only Clifford + T gates.

\begin{minted}[breaklines]{python}
from qrisp import (QuantumBool, QuantumFloat, 
Qubit, invert, control, custom_control)

@custom_control
def mod_adder(a : QuantumFloat, 
              b : QuantumFloat, 
              inpl_adder : callable, 
              modulus : int, 
              ctrl : Qubit = None):
    """
    Performs the modular in-place addition
    b += a
    The adder backend can be specified in form
    of a function "inpl_adder" which performs 
    non-modular addition.
    """
    
    reduction_not_necessary = QuantumBool()
    sign = QuantumBool()

    b = list(b) + [sign[0]]
    
    if ctrl is None:
        inpl_adder(a, b)
    else:
        with control(ctrl):
            inpl_adder(a, b)
            
    with invert():
        inpl_adder(modulus, b)

    cx(sign, reduction_not_necessary)
    
    with control(reduction_not_necessary):
        inpl_adder(modulus, b)
        
    with invert():
        if ctrl is None:
            inpl_adder(a, b)
        else:
            with control(ctrl):
                inpl_adder(a, b)
    
    cx(sign, reduction_not_necessary)
    reduction_not_necessary.flip()
    
    if ctrl is None:
        inpl_adder(a, b)
    else:
        with control(ctrl):
            inpl_adder(a, b)
    
    sign.delete()
    reduction_not_necessary.delete()
\end{minted}
Note the \texttt{custom\_control} decorator which enables an efficient controlled version of this function. To learn more about this feature, please refer to \cite{seidel_2024_qrisp}.\\
Qrisp offers a variety of built-in adder back-end implementations such as the Draper adder \cite{draper_2000}, the Gidney adder \cite{Gidney_2018}, the Cuccaro adder \cite{Cuccaro_2004} or the QCLA introduced by Wang et al. \cite{wang_2023}.

\subsection{Modular adder on arbitrary Montgomery shifts}
\label{sec:mont_form_adder}
As mentioned, instead of reverting back the \mintinline{python}{QuantumModulus} to the standard representation after each quantum-quantum multiplication, it might be more efficient to have the arithmetic interoperable between arbitrary Montgomery shifts. To achieve this, consider two numbers $a, b \in \mathbb{F}_p$ in Montgomery form with the respective Montgomery shift $k, l \in \mathbb{N}$. The sum of both in Montgomery form with shift $l$ is
\begin{align}
    &(a + b) 2^l \text{ mod } p\\
    = &(2^{l-k} a 2^k + b 2^l) \text{ mod } p\\
    = &(\sum_{i = 0}^n \tilde{a}_i 2^{i-k} + b) 2^l \text{ mod } p
\end{align}
Where $\tilde{a}_i$ is the bitstring representing $a 2^k \text{ mod } p$. This bitstring is directly available since $a$ is encoded in Montgomery form with shift $k$. We can now execute the sum as a series of controlled modular additions.
\begin{minted}[breaklines]{python}

def montgomery_addition(a : QuantumModulus, 
                        b : QuantumModulus):
    """
    Performs the modular inplace addition 
    b += a
    where a and b don't need to have the same
    montgomery shift
    """"
    for i in range(len(a)):
        with control(a[i]):
            # .m is the attribute, which 
            # stores the Montgomery shift
            b += 2**(i-a.m)%(a.modulus)
\end{minted}
However, a benchmark of the quantum operations of the two proposed strategies showed that they are comparable, and there is no significant advantage of using the latter instead of the former. We conclude that further research is required to find a definitive answer to this question.

\subsection{Uncomputation}
As with many quantum algorithms, the super-polynomial speed-up of Shor's is fundamentally related to interference phenomena. Interference between several branches can however only occur if they are disentangled completely, implying the EC arithmetic has to act in-place and every temporary value has to be uncomputed. However, in many cases, uncomputation is far from trivial, which posed several challenges when developing the code. In this work, several techniques for uncomputation are leveraged.

\begin{figure*}[htb]
    \centering
    \includegraphics[width = 1\textwidth]{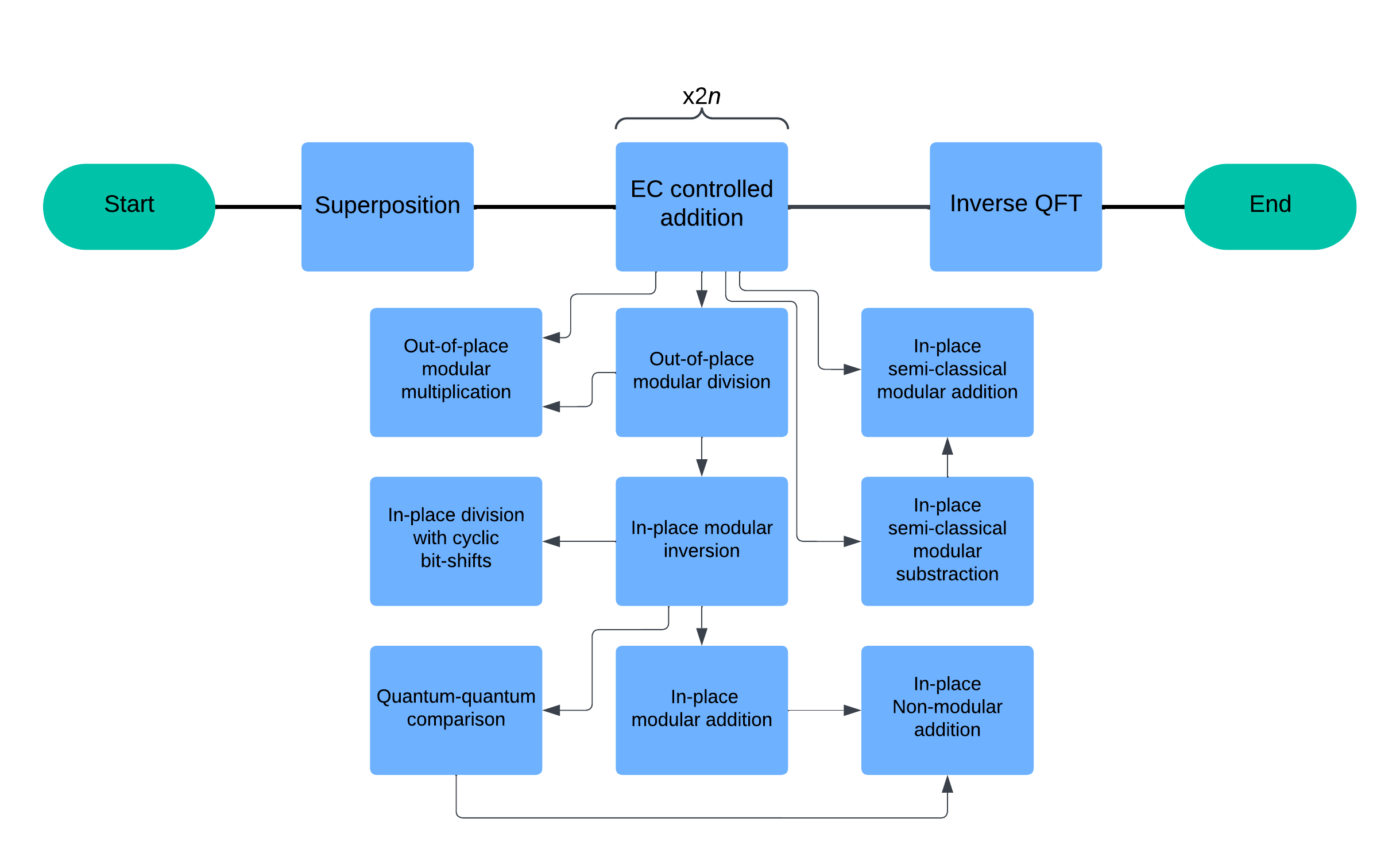}
    \caption{Call graph of Shor's algorithm for solving the ECDLP. We reported here the relevant operations involving \mintinline{python}{QuantumVariable}s. The EC addition is performed $2n$ times, where $n$ is the number of qubit in the control register. }
    \label{fig:callgraph}
\end{figure*}

\begin{itemize}
    \item The \textbf{Unqomp algorithm} \cite{paradis_2021}, which enables an automatic procedure for synthesizing uncomputation. This algorithm has been implemented within Qrisp \cite{Seidel_2023} and can be called using the \texttt{.uncompute} method of the \texttt{QuantumVariable} class. Although this way of uncomputation is especially convenient due to its automation, it has the disadvantage that the values required for the computation need to be accessible\footnote{For further details what accessible means in this context we refer to the Unqomp paper \cite{paradis_2021}.} or otherwise the Unqomp DAG will contain a cycle. This condition is satisfied for some of the uncomputations here, but not for all.

    \item \textbf{Problem specific approaches} to uncomputation. In situations where the Unqomp algorithm fails because the required values for uncomputation are no longer accessible, it is sometimes possible to leverage knowledge about the program state to reverse the computation by using an alternative way for uncomputation. An example of this can be found in Figure \ref{fig:KaliskiIteration}, where the value $a$ is uncomputed in a way that is fundamentally different from its computation. Another interesting example of this is the uncomputation of the "Montgomery garbage" in \cite{rines_2018}.
    
    \item \textbf{Bennett's trick} \cite{Bennett_1973} is a technique for uncomputing the intermediate values of an arbitrary reversible classical function $f$. The procedure can be roughly summarized as follows \begin{enumerate}
        \item Compute the result $y = f(x)$.
        \item Copy $y$ into a result register.
        \item Reverse $f$ by using $y$ and the intermediate values from step 1.
    \end{enumerate}
    While this procedure indeed also applies for classical functions implemented on a quantum computers, it comes with two important drawbacks.
    \begin{itemize}
        \item $f$ has to be evaluated twice (forwards and backwards).
        \item The memory requirements of $f$ grow with the amount of intermediate values required since the occupied space can't be recycled during the execution of $f$.
    \end{itemize}
    Because of the heavy toll on resource requirements, Bennett's trick should in practice therefore be seen as a "last resort" to uncomputation.
\end{itemize}
Within our implementation, we effectively leverage a combination of all these techniques. When Kaliski's algorithm is executed, the $b$ values (see Fig. \ref{fig:KaliskiIteration}) can be essentially uncomputed the same way they have been computed; however, to prevent memory churn\footnote{Memory churn is the compiler overhead that comes with the cost of (de)allocating a lot of variables.}, we manually uncompute (instead of calling \texttt{.uncompute}) to prevent deallocation of $b$, such that the variable can be reused in the next iteration. The $a$ values require the problem-specific approach - for the details we refer to \cite{Gouzien_2023}. The $m$ values seem to allow no direct uncomputation, which is why we had to resort to Bennett's trick \cite{Bennett_1973}. We implement this using the Qrisp in-built \texttt{ConjugationEnvironment} \cite{seidel_2024_qrisp}, which provides a structured interface to achieve this task. Note that the values $a$ and $b$ are uncomputed after each iteration, so the Bennett-related memory overhead is restricted to the values $m$. The call graph of the quantum routines of Shor's algorithm for solving the ECDLP is shown on Fig.\ref{fig:callgraph}.

\section{Circuit benchmark}In this section, we provide the results of benchmarking our implementation, presenting various metrics for each major routine of the quantum algorithm. Figure \ref{fig:benchmarking} shows the maximal number of qubits, T-gate count, CX-gate count, and circuit depth for the major routines, specifically, Kaliski's algorithm, EC addition, controlled EC addition, and the entirety of Shor's algorithm. The parameters used for the elliptic curve are $p=7$, $a=5$, and $b=4$, with the generator of the cyclic subgroup set as $G = (3,2)$. The ancilla register was initialized to $P_0 \in \{(2, 6), (4, 2), (0, 5), (5, 0)\}$, where each initialization corresponds to a simulation of the routine. Since the output values were consistent across these simulations for each routine, they are reported only once. For running the full algorithm, the other classically known point was set to $P = (0,2)$.
The benchmark presented here is fairly important to establish the feasibility of Shor's algorithm to solve ECDLP  on larger scale. Prior to this work, detailed circuit-level implementations were scarce, with only one study reporting qubit counts \cite{jaques_2021}. This scarcity of comparative benchmarks underscores the novelty and potential impact of our contribution to the field.
\paragraph{Qubit count}
The maximal number of qubits required for each routine highlights the resource demands of the implementation in terms of widht of the generated quantum circuit. Kaliski's algorithm utilized $30$ qubits, while EC addition and controlled EC addition required $64$ and $66$ qubits, respectively. Shor's algorithm, encompassing all routines, required $69$ qubits in total. These figures underscore the significant overhead imposed by operations on elliptic curves, especially in the controlled addition routine.
\begin{figure*}[]
\centering
    \begin{tabular}{|c|c|c|c|c|} 
    \hline
    Routine & Qubit count & $\mathrm{T}$-gate count & $\mathrm{CX}$-gate count & T-depth\\
    \hline
    \hline
    Kaliski  & 30 & 2918 & 5651  & 855\\
    \hline
    EC Addition & 64 & 16388 & 37331.5 & 3829\\
    \hline
    EC ctrl-addition  & 66 & 46971 & 112654.5  & 11281\\
    \hline
    Shor's algorithm  & 69 & 93942 & 225246 & 22527\\
    \hline
    \end{tabular}

\caption{Maximal number of qubits, T-gate count, CX-gate count, and depth for the major routines - Kaliski's algorithm, EC addition and controlled EC addition - and the whole Shor's algorithm. The curve employed here is defined by $p=7$, $a=5$, $b=4$. The generator of the cyclic subgroup is $G = (3,2)$, and the ancilla register was initialized to $P_0 \in {(2, 6),(4, 2),(0, 5),(5, 0)}$, each corresponding to a simulation of the routine. For each simulation the values were identical for each routine so they were reported only once.
For running the whole algorithm, the other classically known point was set to $P = (0,2)$.}
\label{fig:benchmarking}
\end{figure*}
\paragraph{Gate count}
The T-gate count and CX-gate count provide insights into the computational cost of the routines. For the CX count, there is an ambiguity. Since we are utilizing Gidney's adder \cite{Gidney_2018} there are CZ gates that are executed conditioned on a measurement result. This measurement returns "True" with a probability of 50\%. In order not to over/underestimate the performance of our implementation we pick the average case, i.e. 0.5 CX gates per classically conditioned CZ.
Kaliski's algorithm had $2918$ T-gates and $5651$ CX-gates, demonstrating a relatively moderate cost compared to the other routines. EC addition required $16388$ T-gates and $37331.5$ CX-gates, while controlled EC addition incurred a notably higher cost with $46971$ T-gates and $112654.5$ CX-gates. Shor's algorithm, combining all components, resulted in a total of $93942$ T-gates and $225246$ CX-gates.
\paragraph{T-depth}
The circuit depth reflects the critical path and parallelizability of the implementation. Kaliski's algorithm exhibited a depth of $855$, while EC addition and controlled EC addition had depths of $3829$ and $11281$, respectively. The complete Shor's algorithm had a depth of $22527$.
\section{Challenges and outlook}
The implementation of Shor's algorithm to solve the ECDLP presents significant challenges both in scalability and optimization. Regarding the former, even the smallest non-trivial instance of the problem requires a substantial amount of resources, as described in the previous section.
While the simulation of larger instances of the problem should remain difficult to handle, as in this difficulty lies the advantage of quantum computation, the compilation of said instances will be tackled through advances in compiler tooling. Even though the foundational building blocks of the elliptic curve logarithm are available within Qrisp, the scalability of the code is still lacking. The problems here arise in a variety of areas. In particular, the limited Python interpreter speed restricts us from compiling the algorithm beyond toy problem scales. Many of these problems are tackled in the MLIR-based compilation pipeline of Qrisp \cite{seidel_2024_qrisp, Ittah_2024, lattner_2020} which will go public in the very near future.

Furthermore, we aim to bring our implementation up to the state-of-the-art by leveraging windowed arithmetic \cite{Gidney_2021}. This feature requires the use of a Quantum Read Only Memory (QROM) to load classical points from a lookup table. This reliance on QROM introduces complexity in the theoretical underpinnings and practical implementation of the algorithm. However, these challenges also highlight areas for innovation, suggesting that advances in quantum computing frameworks and optimization methods could further solidify the robustness and efficiency of ECDLP solutions.
\section{Conclusion}
In this work, we provided and assembled all the necessary components to solve the ECDLP using Shor's algorithm. We detailed the implementation of in-place modular inversion, specifically Kaliski's algorithm, within the Qrisp quantum computing framework. In addition, we describe the implementation of point addition over elliptic curves, its controlled version, and scalar multiplication. Although some routines were developed in previous work, an end-to-end compilable implementation seemed to be missing in the world of open-source software. In fact, many challenges arise when trying to integrate these routines with each other. The current version demonstrates a proof-of-concept for toy model scales. Many of the compiler scalability issues are being tackled in ongoing work. In particular the Jax-based compiler architecture introduced in \cite{seidel_2024_qrisp} is expected to enable scaling to practically relevant scales.
\section{Code availability}
The source code for the implementation of the ECDLP using Shor's algorithm is available at \url{https://github.com/diehoq/quantum-elliptic-curve-logarithm}.
Qrisp is an open-source python framework for high-level programming of quantum computers. The source code is available at \url{https://github.com/eclipse-qrisp/Qrisp}. 

\section{Acknowledgments}
D.P. acknowledges partial funding from SMILS - IC - EPFL. R. S. performed this work in the scope of the PQ-REACT European Union’s Horizon Europe research and innovation programme under grant agreement No. 101119547.

\bibliographystyle{plain}
\bibliography{sources}

\end{document}